\def\sqrtsNN{\mbox{$\sqrt{s_\mathrm{NN}}$}}

\documentclass[12pt]{article}
\usepackage{graphicx}
\begin{document}

\title{The High Transverse Momentum Non-Photonic Electron Measurements in $Au+Au$ collisions at $\sqrtsNN = 200$~GeV at RHIC/STAR}
\maketitle

\author{~~~~~~~~~~~~~Wenqin Xu (for the STAR\footnote{For the full list of STAR authors, see $\\$\texttt{http://www.star.bnl.gov/central/collaboration/authors/authorList.php}. ~~~~~~For the acknowledgements, see$\\$ \texttt{http://drupal.star.bnl.gov/STAR/public/collaboration/acknowledgements}} ~Collaboration)}{
$\\${Department of Physics and Astronomy, University of California, Los Angeles, California 90095, USA
$\\$xwq1985@physics.ucla.edu}
}

\begin{abstract}
We present preliminary results of Non-Photonic Electron (NPE) production, NPE-hadron azimuthal correlation as well as NPE elliptic flow studies at midrapidity using a data set with high statistics and low photon conversion background from $Au+Au$ collisions at $\sqrtsNN = 200$~GeV collected at RHIC in Run 2010.
$\\$Keywords:      Heavy Flavor, Heavy Ion, Energy Loss, Elliptic flow
$\\$Classification: 1320.Fc, 13.20.He,  25.75.Cj
\end{abstract}

%%%%%%%%%%%%%%%%%%%%%%%%%%%%%%%%%%%%%%%%%%%%
%% MAINMATTER
%%%%%%%%%%%%%%%%%%%%%%%%%%%%%%%%%%%%%%%%%%%%

\section{Introduction}
$\\$Due to their large masses, heavy flavor quarks are believed to be unique probes to the strongly coupled QCD matter created in high energy $Au+Au$ collisions at the Relativistic Heavy Ion Collider (RHIC). They are mainly produced in the initial hard scatterings, so the production can be well studied by perturbative-QCD(pQCD)~\cite{Gyulassy95a}. When heavy flavor quarks traverse the QCD medium, they are expected to lose less energy than light flavor quarks due to the dead-cone effect~\cite{Yu01a}, if the dominant energy loss process is gluon radiation. However, recent experimental studies found that the productions of non-photonic electrons (NPE) from
the semi-leptonic decays of charm and bottom hadrons are strongly suppressed at high $p_T$ in central $Au+Au$ collisions relative to that in $p+p$ collisions scaled by the number of binary collisions~\cite{PHENIX10a, STAR07a} and the NPE-hadron azimuthal correlations possibly show a broadening on the away side~\cite{Bertrand09a}. These results suggest some significant contributions from other processes to heavy flavor quark energy loss, which have been simulating various theoretical reconsiderations.

$\\$In order to differentiate model calculations, high precision measurements of NPE production in both $Au+Au$ and $p+p$ collisions, with disentangled charm and bottom
contributions, are wanted. In addition, model calculations with correct dynamics should be able to predict the elliptic flow of heavy flavor quarks at the same time. Simultaneous measurements of NPE production suppression and elliptic flow have better discriminating power on energy loss dynamics than individual measurement alone. The STAR collaboration has recently measured the NPE production with high precision~\cite{STAR11a} and has separated charm and bottom contributions~\cite{STAR10a}, in $p+p$ collisions. In these proceedings, we report the progresses of studying NPE production and elliptic flow in $Au+Au$ collisions. We also report preliminary results of NPE-hadron correlation, which bear the information of the processes of heavy flavor quarks traversing the bulk QCD medium and thus shed light on heavy flavor jet-medium interactions.

\section{Data sets and Analysis method}
$\\$The data sets used here are $Au+Au$ collisions at $\sqrtsNN = 200$GeV collected by the STAR detector during RHIC Run 2010. In these proceedings, we have utilized about 40 Million Minimum Bias triggered events and about 60 Million High-Tower triggered events. The most important detectors used for NPE analyses are the TPC (Time Projection Chamber)~\cite{STAR03a} and the barrel calorimeter system~\cite{STAR03b} including BEMC (Barrel Electromagnetic Calorimeter) and BSMD (Barrel Shower Maximum Detector). In both data sets, electron identification procedures are applied, including, for example, analyzing the shower profiles in BSMD and the energy (in BEMC) over momentum (in TPC) ratios. The calorimeter system largely enhances our ability to identify the electrons but currently constrains the electron kinematic to be $p_T> 2$~GeV/c and $|\eta|< 0.7$. The selected electron sample, inclusive electrons, is composed of mainly NPEs,  photonic electrons and some hadron contamination.  We reconstruct the 2D invariant masses of electron pairs to estimate the contribution of photonic electrons (PE), the electrons from gamma conversions and $\pi^{0}$, $\eta$ Dalitz decays, with reconstruction efficiency $\epsilon$ obtained from simulation tracks embedded into real events, which spans $30 \sim 60\%$ depending on the cuts and the electron $p_T$ (~\cite{STAR10a} has more details on the analysis methods). Then, statistically we have:%We estimated and subtracted the hadron contribution if there are more than 1\% hadrons in the inclusive electron sample. For NPE measurements, for example,  the azimuthal correlation, the following statistical relationship is used:
\begin{equation}
\Delta \phi_{NPE}=\Delta \phi_{inclusive} - (\Delta \phi_{oppo-sign} - \Delta \phi_{same-sign})/\epsilon - \Delta \phi_{hadron}
\end{equation}
%where $\Delta \phi_{NPE}$ is the azimuthal correlation for NPE, and it could be other NPE variables, e.g. production and $v_{2}$. $\Delta \phi_{inclusive}$ is the measurement for inclusive electrons, $ (\Delta \phi_{oppo-sign} - \Delta \phi_{same-sign})/\epsilon$ is the PE contribution and $ \Delta \phi_{hadron}$ is the hadron contribution, which is neglected if there are no more than 1\% hadrons in the inclusive electron sample. 

where $\Delta \phi_{NPE}$ is the azimuthal correlation, but it also could be other NPE variables, e.g. production or $v_{2}$, and the hadron contribution is neglected if there are no more than 1\% hadrons in the inclusive electron sample. 

\begin{figure}[tb]
%\centering
 \begin{tabular}{cc}
 \includegraphics[width=0.5 \linewidth]{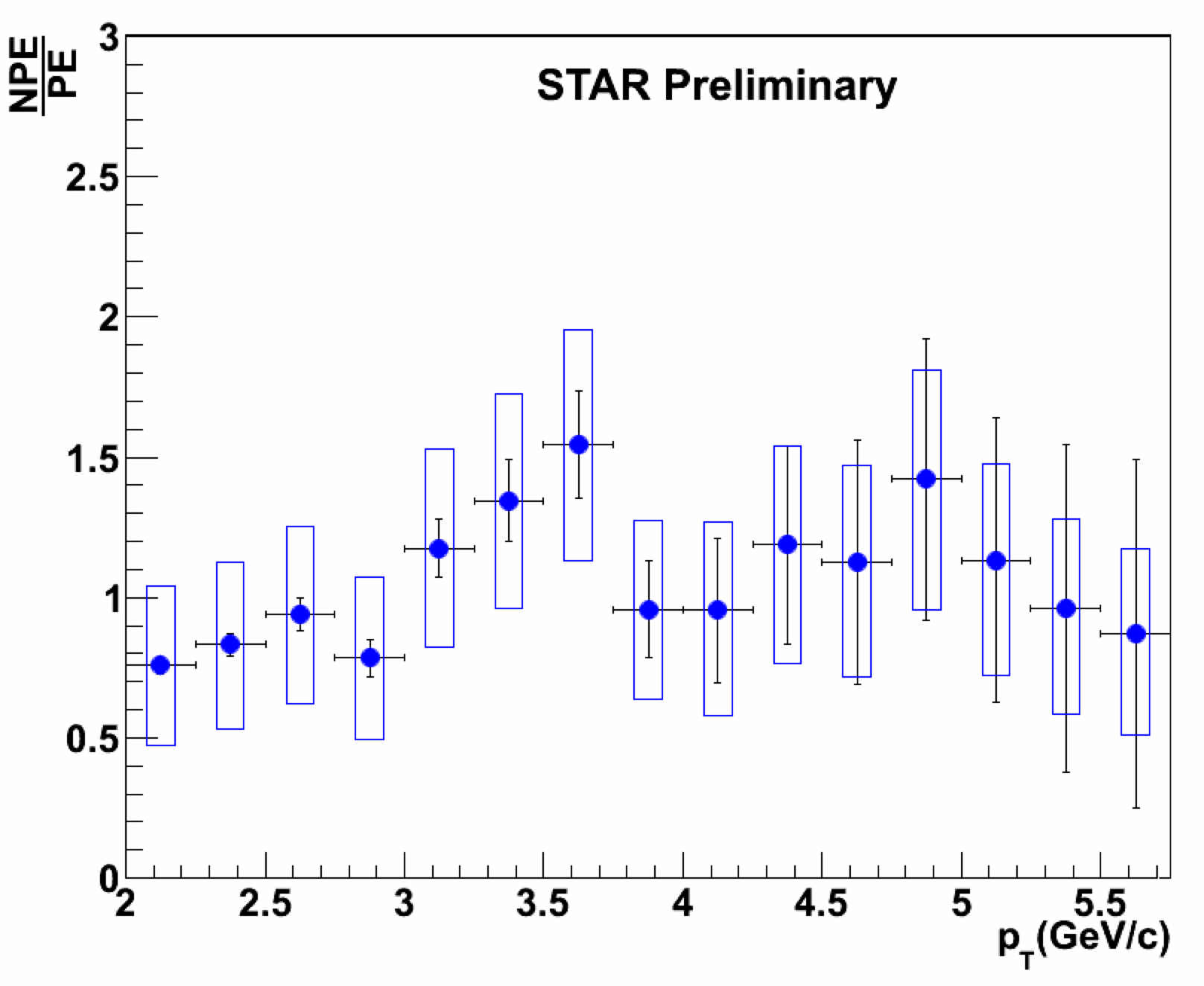}
 \includegraphics[width=0.46 \linewidth]{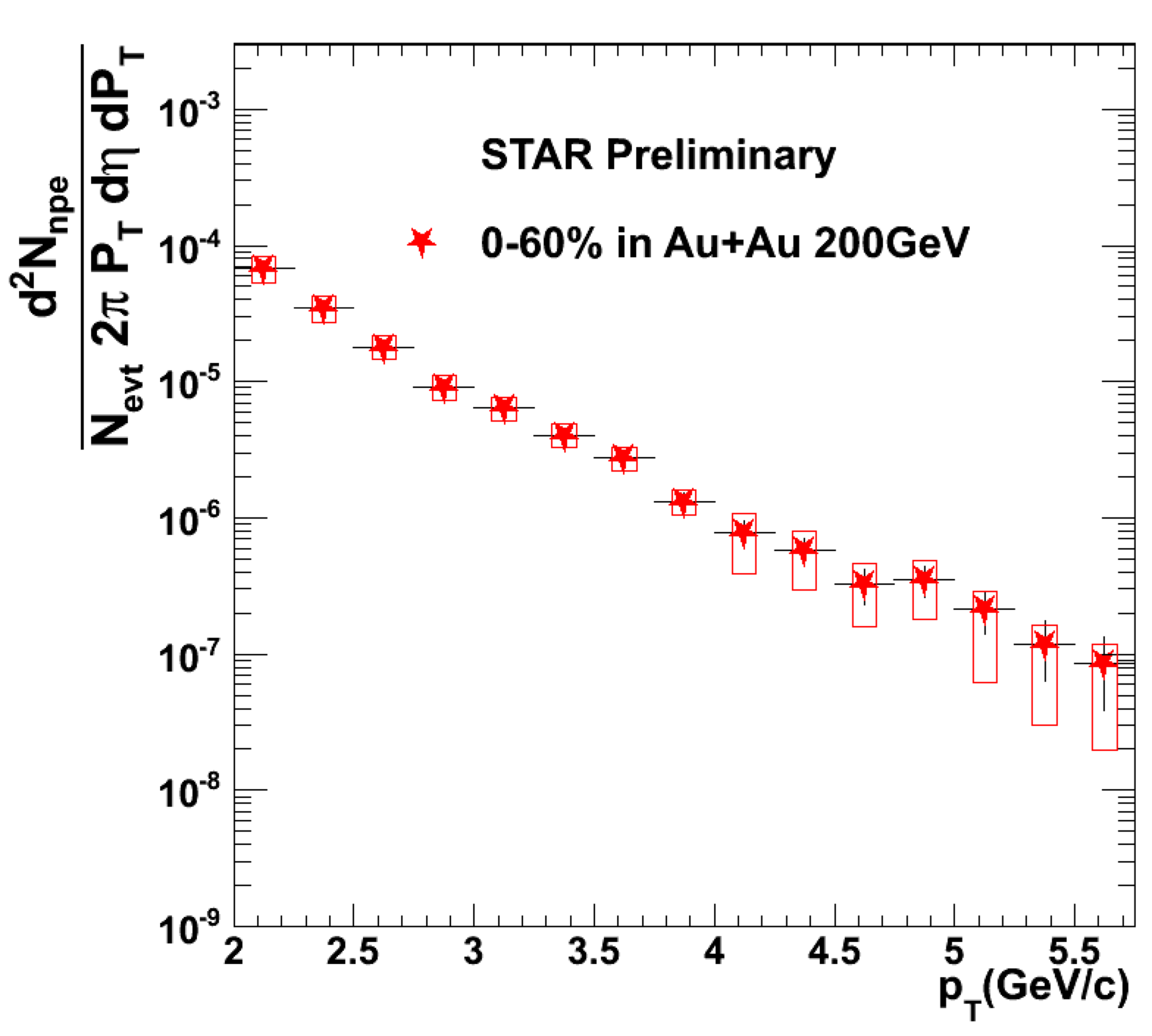}

 \end{tabular}
\caption{Left Panel: Non-Photonic Electron over Photonic Electron ratio as a function of electron $p_T$. Right Panel: The preliminary result of NPE spectrum in $0\sim60\%$ central $Au+Au$ collisions at $\sqrtsNN = 200$~GeV. The systematic errors, shown as boxes, mainly come from the difference between real data and the simulation used in calculating the single electron efficiency and $\pi^{0}$, gamma spectra used in calculating the PE reconstruction efficiency. The uncertainty on electron purity also contributes considerably at high $p_T$.}
\label{fig:spectra}
\end{figure}

\section{Results}
$\\$Figure~\ref{fig:spectra} shows NPE production studies in $0\sim60\%$ central $Au+Au$ collisions based on Minimum Bias triggered events. The left panel demonstrates that the Non-Photonic Electron over Photonic Electron ratio in Run 2010 is around 1, much higher than previous $Au+Au$ runs, providing a cleaner environment for NPE studies. The right panel shows the NPE spectrum with $p_T$ from 2~GeV/c to about 6~GeV/c. For higher $p_T$, High-Tower triggered events will provide larger statistics, ensuring high precision in extended $p_T$ range. In this preliminary result, NPE includes electrons from both open heavy flavor hadrons and heavy quarkonia. In our final results, we will subtract the latter.

\begin{figure}[tb]
%\centering
 \begin{tabular}{cc}
 \includegraphics[height=0.4 \linewidth]{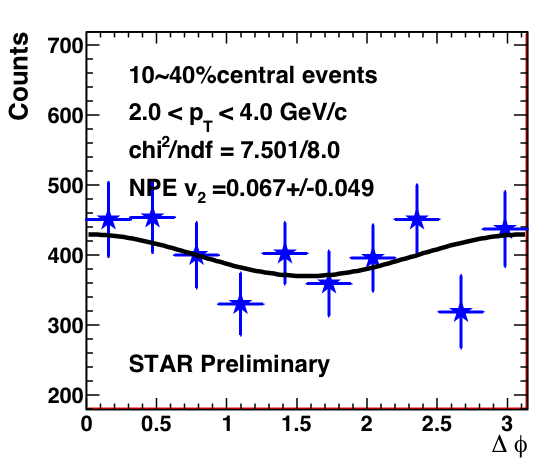}
 \includegraphics[width=0.53 \linewidth]{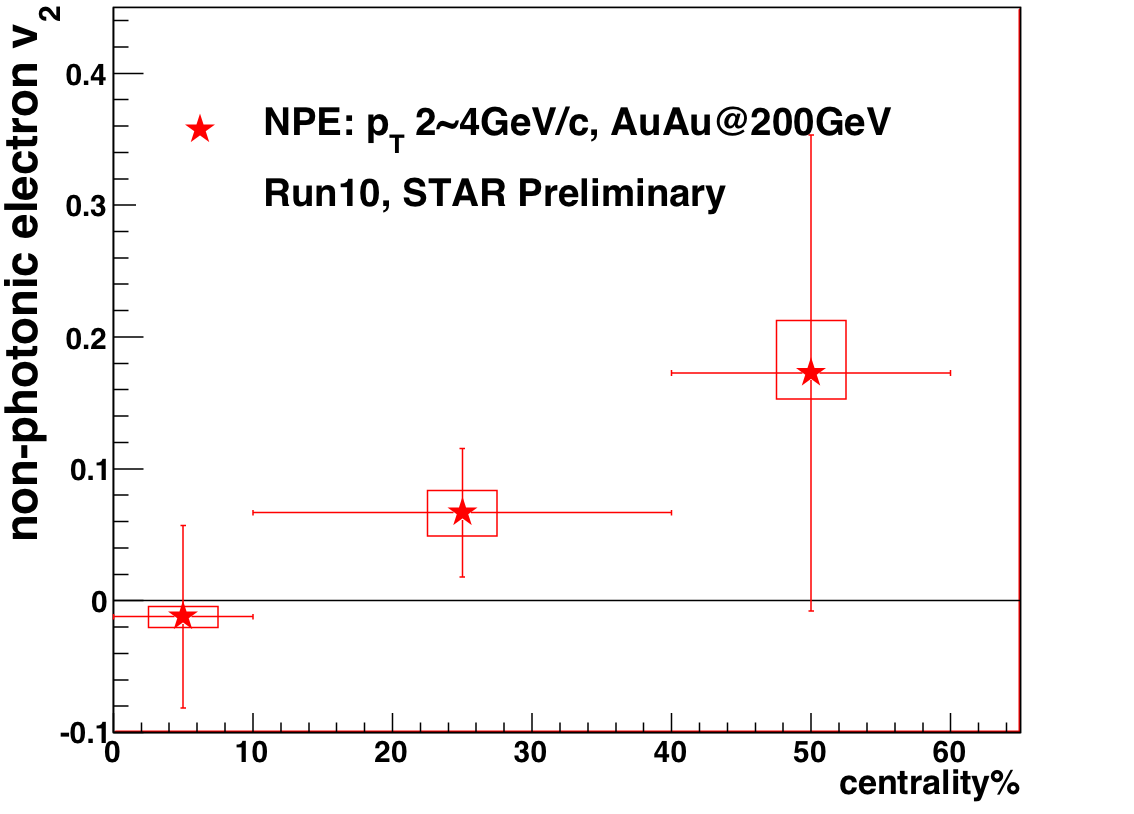}
 \end{tabular}
\caption{Left Panel: Non-Photonic Electron angular azimuthal correlations with the event plane for $10\sim40\%$ central $Au+Au$ collisions at $\sqrtsNN = 200$~GeV. Right Panel: The preliminary results of NPE $v_{2}$ in various centrality bins.}
\label{fig:v2}
\end{figure}

$\\$In Figure~\ref{fig:v2}, the elliptic flow ($v_2$) studies, also based on Minimum Bias triggered events, are shown. We extract $v_2$ by fitting electron angular correlations with the event plane (EP)~\cite{Art98a}. As an example, the $\Delta\phi$ distribution in $10\sim40\%$ central $Au+Au$ collisions for NPE with $2~<p_T^{NPE}<~4$~GeV/c is shown in the left panel. In this middle centrality, the NPE $v_2$ is finite and consistent with our previous study based on Run 2007~\cite{Gang10a}, indicating possible coupling between heavy flavor quarks and the QCD medium. Further conclusions may be drawn only after the statistics error bars get reduced enough as we analyze more statistics. The current systematic uncertainties mainly come from the EP resolution uncertainty due to large centrality bin size and from photonic electron reconstruction efficiency. The differences between different $v_2$ methods are yet to be studied and included. The right panel shows NPE $v_2$ in different centrality bins.%, but the most central and more peripheral bins suffer from very large statistical uncertainties right now.

$\\$In order to ensure that NPE represent well the directions of the parent charm and bottom quarks, we use NPE with $3~<p_T^{NPE}<~6$~GeV/c from High-Tower events to study the NPE-hadron correlation. When we use associated tracks with $0.15~<p_T^{asso}<~0.5$~GeV/c, the tracks that can present the bulk medium the best, the statistical uncertainties are large. With $0.5~<p_T^{asso}<~1.0$~GeV/c, we found there is a NPE-hadron correlation beyond statistical uncertainties on both the near side and away side in $10\sim40\%$ central $Au+Au$ collisions, as shown in Figure 3 in~\cite{Wenqin11a}. If this correlation is also beyond the systematic uncertainties, mainly due to flow modulations which are still under study, then it would indicate heavy flavor tagged jets do interact with the QCD medium. Nevertheless, quantitative measurements yet to emerge and theoretical considerations, such as in~\cite{Jorge09a}, are required to understand the physical picture. 

$\\$In summary, different measurements are sensitive to different aspects of the heavy flavor quark-medium interactions. The centrality dependence of these measurements will reflect properties of the QCD medium and together they provide strong constrains on dynamical models.

\end{document}